\documentclass[journal,10pt]{IEEEtran}

\ifCLASSINFOpdf
\else
\fi

\usepackage[dvips]{color}
\usepackage{graphicx}
\usepackage{subfigure}
\usepackage{amsmath}
\usepackage{mathrsfs}
\usepackage{amsfonts}
\usepackage{breqn}
\usepackage{algorithmic}
\usepackage{algorithm}
\usepackage{amsthm,amsmath,amssymb}
\usepackage{mathrsfs}
\usepackage{float}
\usepackage{xcolor}
\usepackage{tikz}
\usetikzlibrary{shapes.geometric, arrows}

\begin{document}

\title{Securing FC-RIS and UAV Empowered Multiuser Communications Against a Randomly Flying Eavesdropper}
%
% author names and IEEE memberships
% note positions of commas and nonbreaking spaces ( ~ ) LaTeX will not break
% a structure at a ~ so this keeps an author's name from being broken across
% two lines.
% use \thanks{} to gain access to the first footnote area
% a separate \thanks must be used for each paragraph as LaTeX2e's \thanks
% was not built to handle multiple paragraphs
%

\author{Shuying Lin, Yulong Zou,~\IEEEmembership{Senior Member,~IEEE,} Yuhan Jiang, Libao Yang, Zhe Cui, and Le-Nam Tran,~\IEEEmembership{Senior Member,~IEEE}% <-this % stops a space
\thanks{S. Lin, Y. Zou, Y. Jiang, and L. Yang are with the School of Telecommunications and Information Engineering, Nanjing University of Posts and Telecommunications, Nanjing, China. {\emph{(Corresponding author: Yulong Zou.)}}}

\thanks{L.-N. Tran and Z. Cui are with the School of Electrical and Electronic Engineering, University College Dublin, Ireland.}}

% note the % following the last \IEEEmembership and also \thanks -
% these prevent an unwanted space from occurring between the last author name
% and the end of the author line. i.e., if you had this:
%
% \author{....lastname \thanks{...} \thanks{...} }
%                     ^------------^------------^----Do not want these spaces!
%
% a space would be appended to the last name and could cause every name on that
% line to be shifted left slightly. This is one of those "LaTeX things". For
% instance, "\textbf{A} \textbf{B}" will typeset as "A B" not "AB". To get
% "AB" then you have to do: "\textbf{A}\textbf{B}"
% \thanks is no different in this regard, so shield the last } of each \thanks
% that ends a line with a % and do not let a space in before the next \thanks.
% Spaces after \IEEEmembership other than the last one are OK (and needed) as
% you are supposed to have spaces between the names. For what it is worth,
% this is a minor point as most people would not even notice if the said evil
% space somehow managed to creep in.

% The paper headers
%\markboth{}%
{}
% The only time the second header will appear is for the odd numbered pages
% after the title page when using the twoside option.
%
% *** Note that you probably will NOT want to include the author's ***
% *** name in the headers of peer review papers.                   ***
% You can use \ifCLASSOPTIONpeerreview for conditional compilation here if
% you desire.

% If you want to put a publisher's ID mark on the page you can do it like
% this:
%\IEEEpubid{0000--0000/00\$00.00~\copyright~2007 IEEE}
% Remember, if you use this you must call \IEEEpubidadjcol in the second
% column for its text to clear the IEEEpubid mark.

% use for special paper notices
%\IEEEspecialpapernotice{(Invited Paper)}

% make the title area
\maketitle
\vspace{-20cm}
\begin{abstract}
This paper investigates a wireless network consisting of an unmanned aerial vehicle (UAV) base station (BS), a fully-connected reconfigurable intelligent surface (FC-RIS), and multiple users, where the downlink signal can simultaneously be captured by an aerial eavesdropper at a random location. To improve the physical-layer security (PLS) of the considered downlink multiuser communications, we propose the fully-connected reconfigurable intelligent surface aided round-robin scheduling (FCR-RS) and the FC-RIS and ground channel state information (CSI) aided proportional fair scheduling (FCR-GCSI-PFS) schemes. Thereafter, we derive closed-form expressions of the zero secrecy rate probability (ZSRP). Numerical results not only validate the closed-form ZSRP analysis, but also verify that the proposed GCSI-PFS scheme obtains the same performance gain as the full-CSI-aided PFS in FC-RIS-aided communications. Furthermore, optimizing the hovering altitude remarkably enhances the PLS of the FC-RIS and UAV empowered multiuser communications.
\end{abstract}
\vspace{-0.1cm}

\IEEEpeerreviewmaketitle
\vspace{-0.7cm}

\section{Introduction}
The utilization of unmanned aerial vehicles (UAVs) presents a promising solution to enhance wireless capacity and spectrum efficiency \cite{23}, but this leads to potential information leakage to full-space eavesdroppers \cite{14}. To this end, physical-layer designs such as multiuser scheduling have gained significant attention for secrecy enhancement \cite{24}. In multiuser scheduling, a user having better channel state information (CSI) (e.g., a location closer to the RIS or the BS) may consume more time slots than the other users, potentially leading to unfair scheduling. Proportional fair scheduling (PFS) addresses this issue by comparing channel gains normalized by their ergodic value, rather than simply choosing the user with the highest channel gain in conventional scheduling. The fairness of this criterion was proved in [4] to be a simplified version of the original PFS \cite{16} whose scheduling criterion varies in each time slot. Specifically, the authors of \cite{13} investigated UAV-enabled integrated sensing and communications, where the target eavesdropping was prevented by a jammer UAV for securing the communications.
\vspace{-0.1cm}

Despite the security benefits of multiuser scheduling, its superiorities diminish when the ground users are indoor and thus potentially suffer from blocked wireless links or uncontrollable electromagnetic environments \cite{9}. As a remedy, reconfigurable intelligent surface (RIS) offers a lightweight, low-profile, and simple hardware alternative that can be seamlessly integrated into various communication networks, being regarded as a propitious solution that realizes a smart radio environment \cite{21}. The beyond diagonal (BD)-RIS architecture, where each RIS element is connected with the other elements, has been proposed to unlock the beamforming capability of RIS technology \cite{7}. This has inspired variants such as sub-array active RISs and cell-wise single-connected RISs, among which the fully-connected (FC)-RIS is a special case that offers the utmost beamforming gain with all the RIS elements connected with each other. Thus, there is an urgent need to incorporate FC-RISs into UAV-enabled multiuser secure communications. Meanwhile, one of the crucial technical challenges is that the security threats get more severe with additional propagation paths introduced by more flexible connections of FC-RISs.

Moreover, none of the above work considered multiuser scheduling or investigated the hovering altitude of the UAV-BS for FC-RIS-aided communications, especially in terms of improving secrecy performance. Therefore, in this paper, we examine BD-RIS and UAV aided multiuser communications in the presence of a randomly flying eavesdropper. The main contributions of this paper are summarized as follows. First of all, we propose the FC-RIS-aided round-robin scheduling (FCR-RS) scheme, the FC-RIS and GCSI aided proportional fair scheduling (FCR-GCSI-PFS) scheme to improve the security of downlink communications. In addition, we derive closed-form expressions of the zero secrecy rate probability (ZSRP) of the FCR-RS and FCR-GCSI-PFS schemes.
\vspace{-0.4cm}
\section{System Model}
\vspace{-0.2cm}
In this section, we describe the adopted system setting and the corresponding channel model of UAV and FC-RIS aided multiuser secure communications.{\footnote{\emph{Notations}: ${{\bf{a}}^{{T}}}$ and ${{\bf{a}}^{{H}}}$ mean the transpose and conjugate transpose of vectors, respectively. Also, diag(${\boldsymbol{a}}$) denotes a diagonal matrix with its diagonal elements given by ${\boldsymbol{a}}$, E($\cdot$) represents statistical expectation operator, and arg($\cdot$) represents the phase of a complex number. Besides, $n$! represents factorial of a non-negative number $n$, and $\Gamma(\cdot,\cdot)$ represents the upper incomplete gamma function, where $\Gamma(\cdot)=\Gamma(\cdot,0)$ is the gamma function, and $\Gamma(n+1)=n!$.}}
\vspace{-0.4cm}
\subsection{Channel Model}
\vspace{-0.2cm}
\begin{figure}
  \centering
  {\includegraphics[scale=0.16]{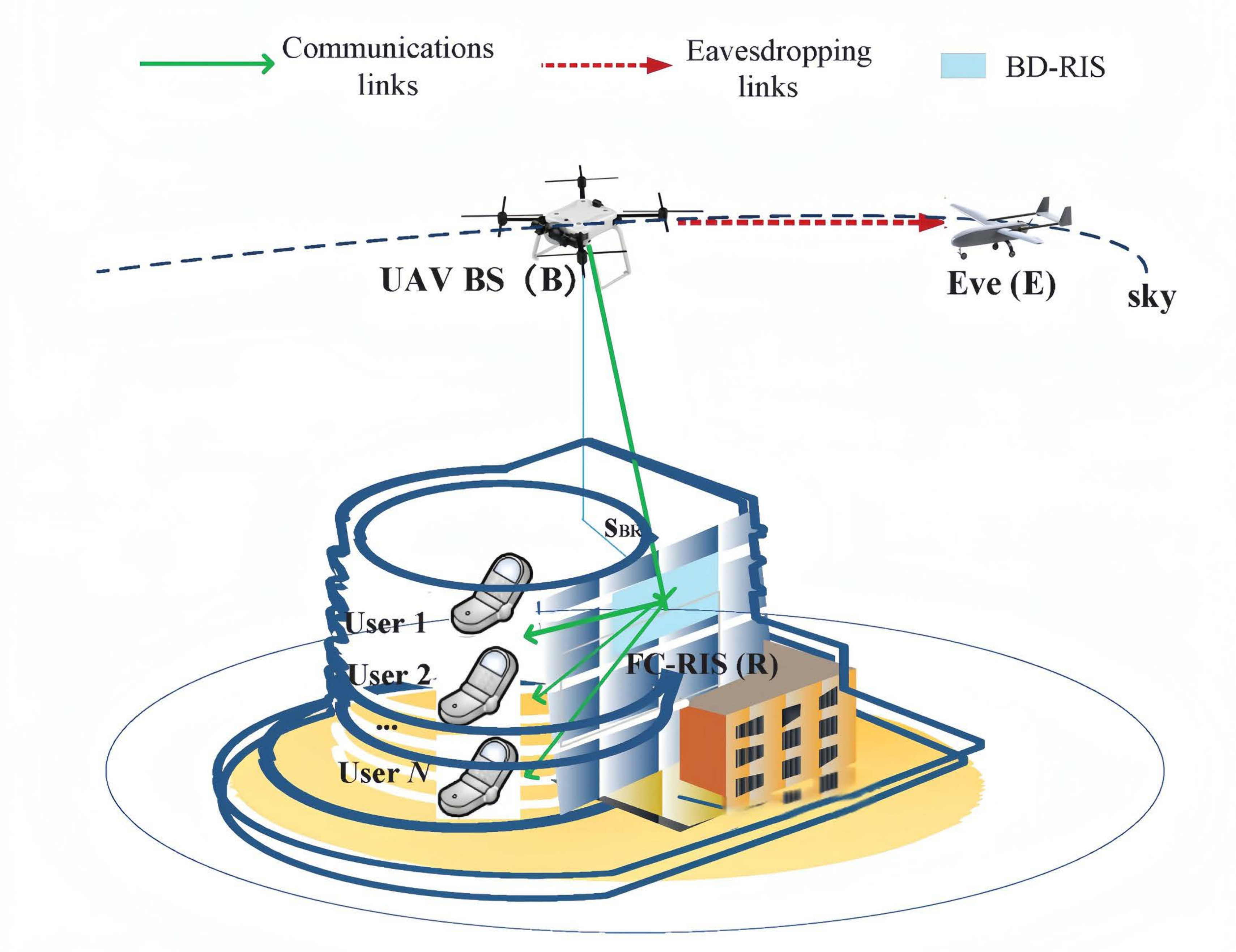}\\
  \caption{The considered FC-RIS and UAV aided multiuser communications network in the presence of an aerial eavesdropper.}\label{Fig1}}
\end{figure}
 We consider a FC-RIS-aided multiuser wireless network consisting of a base station (BS) deployed on a UAV, $N$ ground users denoted by $\mathcal{N}\overset{\Delta}=\{1, 2, \cdots,n, \cdots, N\}$, a BD-RIS with $L$ co-located reflecting elements denoted by $\mathcal{L}\overset{\Delta}=\{1, 2, \cdots ,l, \cdots, L\}$, and a passive eavesdropper (Eve) in the air (e.g., UAV-mounted, etc.), as illustrated in Fig. 1.{\footnote{We consider that each node is equipped with a single antenna. It is worth mentioning that our results can be extended to multi-antenna cases, which is left for future work.}} Without loss of generality, we assume that the BD-RIS locates at the origin of a 3D Cartesian coordinate system, where the location denoted by ${\bf{R}}=(0,0,0)$, the location of the UAV-BS is denoted by ${\bf{B}}=(x_{\text{B}},y_{\text{B}},h_{\text{BR}})$, the location of the Eve is denoted by ${\bf{E}}=(x_{\text{B}}+d_{\text{BE}}\cos\lambda_{\text{E}}\sin\beta_{\text{E}},y_{\text{B}}+d_{\text{BE}}\sin\lambda_{\text{E}}\sin\beta_{\text{E}},h_{\text{BR}}+d_{\text{BE}}\cos\beta_{\text{E}})$, where $d_{\text{BE}}$ denotes the spatial distance between the UAV-BS and the Eve, $\lambda_{\text{E}}\in[0,2\pi)$ and $\beta_{\text{E}}\in[0,\pi)$ are the polar angle and azimuth angle from the Eve to the UAV-BS \cite{25}, respectively. Besides, the users are positioned at fixed locations around the central origin (e.g., the $n$-th location is denoted by ${\bf{U}}_n=(x_n,y_n,0)$).{\footnote{${\bf{U}}_n,\;\forall n\in\mathcal{N}$ can be either available from the global database or with standard positioning techniques (e.g., Global Positioning System) \cite{1}.}} Among the user set $\mathcal{N}$, a particular user indexed by $n$ is opportunistically selected to communicate with the aerial BS according to the channel quality of ground links, which will not change as rapidly as the links related to the high-mobility UAV. The CSI knowledge can be available through RIS-aided channel estimations \cite{2}, where the UAV-BS gets access to a database that comprises of all related CSI, or exchanges the control information with each user. Taking into account the uncertain location of the Eve, we assume it is uniformly-distributed within a hemisphere with a radius $R$ surrounding the UAV-BS, $d_{\text{BE}}\le R$. Besides, we assume quasi-static block-fading channels for a link related to the FC-RIS, and each channel coefficient is modeled as
 \vspace{-0.1cm}
\begin{equation}\label{equa(1)}
 \mathbf { h }_{k}=\sqrt{\mathbb{G}_{k}}\mathbf {g}_{k},
 \end{equation}
 where the subscript $k\in\{{\text{R}n},{\text{BR}}\}$ represents links from the FC-RIS to the $n$-th user and from the UAV-BS to the FC-RIS, $\mathbf { h }_{k}\in{{\mathbb{C}}^L}$, $\mathbf { g }_{k}\in{{\mathbb{C}}^L}$, and ${\mathbb{G}}_k={\mathbb{G}}_0d_k^{-\alpha_k}$ stands for the large-scale fading, where ${\mathbb{G}}_0$ denotes the reference fading loss at the reference distance, $\alpha_k$ and $d_k$ denote the path loss exponents and the respective distances expressed as $d_{\text{R}n}=||{\bf{R}}-{\bf{U}}_n||,\quad d_{\text{BR}}=||{\bf{B}}-{\bf{R}}||,\quad d_{\text{BE}}=||{\bf{B}}-{\bf{E}}||$.
  Following \cite{3} and \cite{4}, we consider an altitude-dependent path loss exponent for modelling an air-to-ground links, which typically varies over the density of obstacles encountered. Hence, the path loss exponent $\alpha_{\text{BR}}$ is characterized as below{\footnote{The approximations $ P^{\text{LoS}}(\frac{\pi}{2})=1$ and $ P^{\text{LoS}}(0)=\frac{1}{1+a_2{\rm{exp}}({a_2b_2})}$ are substituted into (2) to determine the coefficients $a_1$ and $b_1$ as $a_1=(\alpha({\pi}/2)-\alpha(0))\frac{1+a_2{\rm{exp}}(a_2b_2)}{a_2{\rm{exp}}(a_2b_2)}$ and $b_1=\alpha(0)-\frac{a_1}{1+a_2{\rm{exp}}(a_2b_2)}$, where $\alpha({\pi}/2)$ and $\alpha(0)$ can be obtained by measurements, whose adopted values of $\alpha({\pi}/2)$ and $\alpha(0)$ will be specified in the simulation section, and $\alpha_{\text{R}n}=\alpha(0),\;\forall n\in\mathcal{N}$.}}
 \vspace{-0.1cm}
  \begin{equation}\label{equa(2)}
\alpha_{\text{BR}}=a_1P_{\text{BR}}^{\text{LoS}}+b_1,
 \end{equation}
 where $P_{\text{BR}}^{\text{LoS}}$ is the probability of having a line-of-sight (LoS) link, which is given by
 \vspace{-0.1cm}
 \begin{equation}\label{equa(3)}
 P^{\text{LoS}}(\theta_{\text{BR}})=\frac{1}{1+a_2{\rm{exp}}(-b_2(\theta_{\text{BR}}-a_2))},
 \end{equation}
 where $\theta_{\text{BR}}$ is the elevation angle, i.e., $\theta_{\text{BR}}=\arctan(\frac{h_{\text{BR}}}{d_{\text{BR}}})$, $d_{\text{BR}}=\sqrt{x_{\text{B}}^2+y_{\text{B}}^2}$, $h_{\text{BR}}$ is the flying altitude of the UAV, and $d_{\text{BR}}$ is the projected horizontal distance between the BS and the RlS. Besides, $a_2>0$ and $b_2>0$ are positive parameters depending on the environment \cite{5}, \cite{6}.

 Taking into account the randomness of wireless channels, we adopt time-varying small-scale fading $\mathbf {g}_{k}$ to model the air-to-ground channel and the ground channels. Considering the applicability in various scenarios with 3D full-space coverage of the nodes, we employ a more general Nakagami-$m$ fading model,{\footnote{With such a general channel model, the analysis can be readily extended to more complicated scenarios such as the satellite-UAV-terrestrial networks with negligible differences. Restricting the fading parameter $m$ to integer values (often 2-6) simplifies mathematical analysis and enables the derivation of closed-form expressions.}} where the single-link channel gain $|{g}_{k}|$ is specified as a unit-variance variable, i.e., $\scriptsize{{\mathbf { g }}_{\text{R}n}=\frac{||{\mathbf { h }}_{\text{R}n}||}{\text{E}(||{\mathbf { h }}_{\text{R}n}||)}}$, $\scriptsize{{\mathbf { g }}_{\text{BR}}=\frac{||{\mathbf { h }}_{\text{BR}}||}{\text{E}(||{\mathbf { h }}_{\text{BR}}||)}}$, with a distribution characterized by
 \vspace{-0.1cm}
  \begin{equation}\label{equa(4)}
\Pr(|{g}_{k}|\le g)=1-\frac{\Gamma(m_k,m_k\sqrt{g})}{\Gamma(m_k)},\quad\forall k\in\{{\text{R}n},{\text{BR}}\},\;\forall n\in\mathcal{N},
 \end{equation}
 where $\frac{\gamma\left(m_kL, m_k\sqrt{g}\right)}{\Gamma(m_k)}=1-e^{-m_k\sqrt{g}} \sum_{t=0}^{m_k-1}{\frac{1}{\Gamma(t+1)}}$ $\left(m_k\sqrt{g}\right)^t$. Analogous to (1), we note $h_{\text{BE}}=\sqrt{{\mathbb{G}}_0}d_{\text{BE}}^{-\alpha_{\text{BE}}/2}g_{\text{BE}}$. However, the link between the UAV BS and the aerial Eve (air-to-air channels) is assumed to be LoS \cite{14}, \cite{13}, i.e., $\text{E}({g}_{\text{BE}})=g_{\text{BE}}\in{{\mathbb{C}}^1}$.

As an extension of conventional RISs \cite{7}, BD-RISs have recently been proposed with their novel architectures of a multi-port scattering parameter network with each RIS element being its input/output, connected to one another in groups. Notably, we focus on a FC architecture when there is just one connected group including elements $l\in\mathcal{L}$. Specifically, the scattering coefficient matrix $\boldsymbol{\Theta}_n$ of the FC-RIS is not limited to be diagonal and is expressed as the following complex symmetric unitary matrix
 \vspace{-0.1cm}
  \begin{equation}\label{equa(5)}
\boldsymbol{\Theta}_n^H\boldsymbol{\Theta}_n=\mathbf{I}_L,\;\;\boldsymbol{\Theta}_n=\boldsymbol{\Theta}_n^T,\quad \forall n\in\mathcal{N},
\end{equation}
 which extends the unit-modulus constraint concerning conventional diagonal RISs.
\vspace{-0.3cm}
\subsection{Signal Model}
 It can be understood from Fig. 1 that the UAV multiuser communications of interest is purely established by the scattering of FC-RIS installed to the building, because the signal transmissions from the UAV-BS to the indoor users located in the building may be seriously blocked. In this case, we focus on the cascaded links through the FC-RIS without the direct links. When the aerial BS starts to transmit a joint signal $x_{\text{B}}\in{\mathbb{C}}$ at a power of $P_{\text{B}}$, the signal received at the $n$-th user can be written as
 \vspace{-0.1cm}
\begin{equation}\label{equa(6)}
 y_{\text{B}n}=\sqrt{P_{\text{B}}}\left({\bf{h}}^H_{\text{R}n}{\boldsymbol{\Theta}}_{n} {\bf{h}}_{\text{BR}}\right) x_{\text{B}}+n_n,
\end{equation}
 where ${\bf{h}}_{\text{BR}}\in{{\mathbb{C}}^{L\times1}}$ and ${\bf{h}}_{\text{R}n}\in{{\mathbb{C}}^{L\times1}}$ represent channel coefficients of RIS-BS transmissions and the $n$-th user to the RIS, respectively. Besides, $n_n$ is an additive white Gaussian noise (AWGN) at the $n$-th user, which is with zero mean and $N_0$ variance. Upon the Shannon theorem, the instantaneous capacities of the main links are given by
 \vspace{-0.1cm}
 \begin{equation}\label{equa(7)}
  C_{\text{B}n}= {\log _2}\left(1 + \frac{P_ {\text{\text{B}}}}{N_0}|{\bf{h}}^H_{\text{R}n}{\boldsymbol{\Theta}}_{n} {\bf{h}}_{\text{BR}}|^2\right).
\end{equation}
\vspace{-0.1cm}

 Due to the broadcast nature of wireless propagation, the BS-user transmissions can also be overheard by an Eve. Then, the received signal at the Eve is expressed as
 \vspace{-0.1cm}
\begin{equation}\label{equa(8)}
y_{\text{B}\text{E}}=\sqrt{P_{\text{B}}}{h}_{\text{BE}} x_{\text{B}}+n_{\text{E}},
\end{equation}
 where ${h}_{\text{BE}}$ is the channel coefficient of BS-Eve transmissions, and $n_{\text{E}}$ is an AWGN at the Eve. Hence, the instantaneous capacity of the wiretap link is given by
 \vspace{-0.1cm}
\begin{equation}\label{equa(9)}
 C_{\text{B}\text{E}} = {\log _2}\left(1 +  \frac{P_ {\text{\text{B}}}}{N_0}|{h}_{\text{BE}}|^2\right),
\end{equation}
 which completes the system model of FC-RIS aided multiuser UAV communications.
\vspace{-0.5cm}
\section{Proposed Multiuser Scheduling Schemes for FC-RIS-Assisted UAV Communications and ZSRP Analysis}
 In this section, we propose the fully-connected RIS aided round-robin scheduling (FCR-RS) scheme, the FC-RIS and ground CSI aided proportional fair scheduling (FCR-GCSI-PFS) scheme, and then carry out ZSRP analysis in closed forms.
\vspace{-0.3cm}
\subsection{GCSI-Aided Scheduling and Phase Design with a FC-RIS}
 In this section, we propose a GCSI-based multiuser scheduling framework which is specially designed for the FC-RIS-assisted communications. Specifically, more CSI uncertainty may be brought to air-to-ground links by the jittering effect of the UAV-BS \cite{11}. If the ground users are scheduled according to the CSI of their links related to the ground RIS, the information exchange tend to be of lower complexity. It is worth mentioning that the considered multiuser scheduling framework relies on a co-phase alignment of the FC-RIS \cite{12}. By using the eigenvalues decomposition and utilizing the property given by (5), $\boldsymbol{\Theta}_n$ is further given by
 \vspace{-0.1cm}
 \begin{equation}\label{equa(10)}
 \boldsymbol{\Theta}_n=\mathbf{V}\mathbf{D}_n \mathbf{V}^T,\quad\forall n,
\end{equation}
 where $\mathbf{D}_n={\rm diag} ([e^{-j\phi_{n,1}},\cdots ,e^{-j\phi_{n,l}}, \cdots , e^{-j\phi_{n,L}}])$ is a diagonal matrix, wherein $\phi_{n,\;l}\in[0,2\pi)$ denotes the phase shift induced by the $l$-th FC-RIS element. Note that the channel gain given in (7) can be expanded using the Cauchy-Schwarz inequality as
 \vspace{-0.1cm}
 \begin{equation}\label{equa(11)}
|\hat{\mathbf{h}}_{\text{R}n}^H \mathbf{D}_n\hat{\mathbf{h}}_{\text{BR}}|^2\leq\|\hat{\mathbf{h}}_{\text{BR}}^H\|^2\|\mathbf{D}_n\hat{\mathbf{h}}_{\text{R}n}\|^2,
\end{equation}
 where $\hat{\mathbf{h}}_{\text{BR}}^H=\mathbf{h}_{\text{BR}}^H \mathbf{V}$, $\hat{\mathbf{h}}_{\text{R}n}=\mathbf{V}^T \mathbf{h}_{\text{R}n}$, and the equality is achieved when the phase shifts introduced by the FC-RIS align with the cascaded channel coefficients as to maximize the average gain. Thus, the desired phase shifts are given by
 \vspace{-0.1cm}
 \begin{equation}\label{equa(14)}
\phi_{n,l}=-\arg \left(\left[{\mathbf{h}}_{\text{BR}}\mathbf{V}\right]_l\right)-\arg ([\mathbf{V}^T {\mathbf{h}}_{\text{R}n}]_l),\quad\forall l,
\end{equation}
 where $\mathbf{V}$ is readily obtained [20, Algorithm 1] for single-input-single-output communications. With these optimal phase shifts, the equality in (11) holds. Then, we denote the average gain of the FC-RIS-aided B-$n$ channels as $\scriptsize{Z_n=\|\hat{\mathbf{h}}_{\text{BR}}^H\|^2\|\mathbf{D}_n\hat{\mathbf{h}}_{\text{R}n}\|^2}$$=\sigma^2_1\sigma^2_2\|\hat{\mathbf{g}}_{\text{BR}}^H\|^2\|\mathbf{D}_n\hat{\mathbf{g}}_{\text{R}n}\|^2$ for notational convenience. As such, the multiuser scheduling within $n\in\mathcal{N}$ can be equivalently presented as the proposed GCSI-based multiuser scheduling. Notably, if this method is adopted in conventional-RIS aided communications, the selection criterion becomes biased since each element of RIS corresponds to a different random link between the RIS and the UAV BS.

 By considering that the location of eavesdroppers is usually unknown, this randomness is characterized by a uniformly-distributed location in the UAV-BS-centric 3D space with a radius of $R$, which is the maximum distance between the UAV-BS and the Eve. Then, the probability density function (PDF) of $d_{\text{BE}}$ can be expressed as \cite{26}
 \vspace{-0.1cm}
\begin{equation}\label{equa(13)}
f_{d_{\text{BE}}}(\psi)=\frac{3\psi^2}{R^3},\quad 0\le\psi\le R.
 \end{equation}
 For notational convenience, $k=1$ means ${\text{R}n}$, the path from the FC-RIS to the $n$-th user and $k=2$ means ${\text{BR}}$, the path from the UAV BS to the FC-RIS. As can be observed from (4), we assume independently and identically distributed (i.i.d.) Nakagami-$m$ fading channels from different reflecting elements of the FC-RIS, i.e., $\scriptsize{{\mathbf { h }}_{\text{R}n}\sim\mathcal{G}\left({\mathbf{m_1}},\frac{{\boldsymbol{\sigma}}_1^2}{\mathbf{m_1}}\right)\in{{\mathbb{C}}^L}}$ and $\scriptsize{{\mathbf { h }}_{\text{BR}}\sim\mathcal{G}\left({\mathbf{m_2}},\frac{{\boldsymbol{\sigma}}_2^2}{\mathbf{m_2}}\right)\in{{\mathbb{C}}^L}}$, respectively, where $\mathcal{G}$ denotes the gamma distribution. To be specific, the corresponding parameters of the channel fadings are written as
 \vspace{-0.1cm}
 \begin{equation}\label{equa(15)}
 {\boldsymbol{\sigma}}^2_1={\mathbb{G}}_0d_1^{-\alpha_1}\cdot{\mathbf{1}},\; {\boldsymbol{\sigma}}^2_2={\mathbb{G}}_0d_2^{-\alpha_2}\cdot{\mathbf{1}},\; {\mathbf{m_1}}=m_1\cdot{\mathbf{1}},\;{\mathbf{m_2}}=m_2\cdot{\mathbf{1}}.
 \end{equation}
By letting $S_n=\|\mathbf{D}_n\hat{\mathbf{g}}_{\text{R}n}\|^2$ and $W=\|\hat{\mathbf{g}}_{\text{BR}}^H\|^2$, the cumulative density function (CDF) of $S_n$ and the PDF of $W$ are respectively given by
 \vspace{-0.1cm}
  \begin{equation}\label{equa(16)}
\Pr(S_n\le s)=1-\frac{\Gamma(m_1L,m_1s)}{\Gamma(m_1L)},
 \end{equation}
\vspace{-0.1cm}and
 \vspace{-0.1cm}
 \begin{equation}\label{equa(17)}
{f_{{W}}}(w) = \frac{w^{m_2L - 1}{m_2^{m_2L}e^{ - m_2w}}}{\Gamma(m_2L)}.
\end{equation}
\vspace{-0.9cm}
 \subsection{ZSRP of the FCR-RS Scheme}
 In this section, we present the FCR-RS scheme where the downlink communications of each user takes place orderly in a slot-by-slot basis. In the FCR-RS scheme, the overall zero secrecy rate probability (ZSRP) of the multiuser system can be written as
 \vspace{-0.1cm}
 \begin{equation}\label{equa(18)}
 {P_{\text{ZSR, FCR-RS}}} = \frac{1}{N} \sum\limits_{n = 1}^N {P_{\text{ZSR}, n}}.
\end{equation}
 By following the physical layer security literatures \cite{10}, the ZSRP of the $n$-th user can be expressed as
 \vspace{-0.1cm}
\begin{equation}\label{equa(10)}
 P_{\text{ZSR},\; n} = \Pr(C_{\text{B}n}<C_{\text{B}\text{E}}).
\end{equation}
\linespread{1.5}
  {Substituting (7) and (9) into (18) and letting $\alpha_{\text{BE}}=2$ in $\scriptsize{\sigma_{\text{BE}}^2={\mathbb{G}}_0d_{\text{BE}}^{-\alpha_{\text{BE}}}}$ result in}
 \vspace{-0.1cm}
\begin{equation}\label{equa(20)}
 P_{\text{ZSR}, n}=\Pr\left(Z_n<\frac{{\mathbb{G}}_0}{d_{\text{BE}}^2} \right),
\end{equation}
\linespread{1.5}
 {where $\scriptsize{\gamma_{\text{B}}=P_ {\text{B}}/N_0}$, $\scriptsize{Z_n=S_nW}$. Now using (15),  (16), and [22, Eq. (3.471.9)], the CDF of $Z_n$ is expressed as}
 \vspace{-0.1cm}
\begin{scriptsize}
\begin{equation}\label{equa(21)}
\begin{aligned}
&F_{Z_n}(z)=\Pr\left(\sigma^2_1S_n<\frac{z}{\sigma^2_2W}\right)\\
&= 1-\frac{2}{\Gamma(m_2L)N} \sum\limits_{n = 1}^N {} \sum_{t=0}^{m_1L-1}{\frac{1}{\Gamma(t+1)}\left(\frac{z}{\sigma_1^2\sigma_2^2}\right)^{\frac{m_2L+t}{2}}}\mathcal{K}_{m_2L-t}\left( \frac{2\sqrt{z}}{\sigma_1\sigma_2}\right),
\end{aligned}
\end{equation}
\end{scriptsize}
 where $\mathcal{K}_v(\cdot)$ is the $v$-th order Bessel function of the second kind [14, Eq. (8.432.6)], $\scriptsize{\sigma_1^2}$ and $\scriptsize{\sigma_2^2}$ are given by (14), and the superscript ``1" denotes R$n$, $\scriptsize{\forall n\in\mathcal{N}}$. To proceed, the ZSRP of the $n$-th user can be rewritten as
 \vspace{-0.1cm}
\begin{equation}\label{equa(22)}
 P_{\text{ZSR}, n}=\int_0^{R} f_{d_{\text{BE}}}(\psi) F_{Z_n}\left(\frac{{\mathbb{G}}_0}{\psi^{2}}\right){\mathrm{d}}\psi.
\end{equation}
By plugging (13) and (20) into (21), we obtain the closed-form ZSRP of the FCR-RS scheme as (22) at the top of next page, where $G _ { m,n } ^ { p,q }(\cdot)$ is the Meijer G-function [14, Eq. (9.301)], $\scriptsize{\vartheta=\frac{2\sqrt{{\mathbb{G}}_0}}{\sigma_1\sigma_2R}}$, $\scriptsize{\mu=m_2L+t-4}$, $\scriptsize{\varrho=m_2L-t}$, and [22, Eq. (6.592-4)] is used.
 \vspace{-0.1cm}
 \begin{table*}
\begin{equation}\label{equa(23)}
\begin{array}{l}
\begin{aligned}
&{P_{\text{ZSR, FCR-RS}}}= 1-\frac{1}{\Gamma(m_2L)N} \sum\limits_{n = 1}^N {} \sum_{t=0}^{m_1L-1} \frac{3}{\Gamma(t+1)R}\left(\frac{\vartheta}{2}\right)^{m_2L+t}\frac{4^{\mu}}{{\vartheta}^{2\mu}}G _ { 1,3 } ^ { 3,0 } \left( \frac { \vartheta^2 } { 4 }\bigg| \begin{array} { c } 0 \\ - 1  , \frac{\varrho}{2}+\mu,  -\frac{\varrho}{2}+\mu \end{array} \right).
 \end{aligned}
\end{array}
\end{equation}
\hrule
\end{table*}
\vspace{-0.5cm}
\subsection{ZSRP of the FCR-GCSI-PFS Scheme}
 In this section, we propose the FCR-GCSI-PFS scheme to improve the secrecy rate while guaranteeing the scheduling fairness between users of various locations. To avoid always selecting the user closer to the FC-RIS, the selection criterion of the PFS is given by
 \vspace{-0.1cm}
 \begin{equation}\label{equa(23)}
\begin{aligned}
{o}={{\rm{arg}}\mathop{\rm{max}}\limits_{n\in\mathcal{N}}}\left\{S_n\right\},
 \end{aligned}
\end{equation}
\vspace{-0.1cm}
 where $o$ is the index of the scheduled user. With this scheme adopted, the overall ZSRP of the multiuser system can be written as
 \vspace{-0.1cm}
 \begin{equation}\label{equa(24)}
 {P_{\text{ZSR, FCR-GCSI-MS}}} = P_{\text{ZSR},\;o} =\Pr\left(\sigma_{\text{R}o}^2S_o<\frac{z}{\sigma^2_2W}\right).
\end{equation}
\vspace{-0.1cm}
 By referring to (23), the CDF of $S_o$ can be further derived as
 \vspace{-0.1cm}
 \begin{equation}\label{equa(25)}
{\rm F}_{S_o}(s)=\prod\limits_{n\in\mathcal{N}}{\rm F}_{S_n}(s),
\end{equation}
\vspace{-0.1cm}
 from which and by capitalizing on the generalized multinomial theorem, we represent the overall small fading of the main link as
 \vspace{-0.1cm}
\begin{equation}\label{equa(26)}
\begin{aligned}
{\rm F}_{Z_o}(s)=&1- \frac{1}{N}\sum\limits_{n = 1}^N {}\sum\limits_{q = 1}^{{2^N} - 1}\frac{2{( - 1)}^{|{J_{q}}|}}{\Gamma(m_2L)} \sum\limits_{n\in J_q}\sum_{\zeta} {A_1} s^{\frac{m_2L+B_1}{2}} \\&\times \left(\frac{|{J_{q}}|\sigma_2^2}{\sigma_1^2}\right)^{\frac{m_2L+B_1}{2}} \mathcal{K}_{m_2L-B_1}\left(2 \sqrt{\frac{s}{\sigma_1^2\sigma_2^2}}\right),
\end{aligned}
\end{equation}
 \vspace{-0.1cm}
 where ${J_{q}}$ represents the $q$-th non-empty subcollection of the user set $\mathcal{N}$, $|{J_{q}}|$ denote the cardinality of the set $J_{q}$, the set $\zeta=\{\left( n _ { 1 } , n _ { 2 } , \ldots , n _ { m_1L } \right) |\sum _ { p = 1 } ^ {m_1L } n _ { p } = |{J_{q}}|\}$,  $A_1=\frac {\prod_ { k = 1 } ^ {m_1L}{ \frac{1}{(\Gamma(k)) ^ {  n _ { k } }}}}{\Gamma(B_1+1) \prod_ { p = 1 } ^ { m_1L } \Gamma(n _ { p } +1) } $, $B_1=\sum _ { p = 1 } ^ {m_1L } {n _ { p } (p - 1)}$. Analogous to (21), and by combining (13) with (26), the closed-form ZSRP expression of the FCR-GCSI-PFS scheme is given by (27) at top of next page, where $\scriptsize{\rho=m_2L+B_1-4}$, and $\scriptsize{\frac{1}{N}}$ is the duty circle, arising from the fairness of the PFS.
 \vspace{-0.1cm}
 \begin{table*}
\begin{scriptsize}
 \begin{equation}\label{equa(27)}
\begin{array}{l}
\begin{aligned}
 &{P_{\text{ZSR, FCR-GCSI-MS}}}=1- \sum\limits_{n = 1}^N {}\sum\limits_{q = 1}^{{2^N} - 1}\frac{{( - 1)}^{|{J_{q}}|}}{\Gamma(m_2L)N} \sum\limits_{n\in J_q}\sum_{\zeta} { \frac{3A_1 \left({|{J_{q}}|\sigma_2^2}\right)^{\frac{m_2L+B_1}{2}}}{\Gamma(t+1)R}} \left(\frac{\vartheta}{2}\right)^{m_2L+B_1}\frac{4^{\rho}}{{\vartheta}^{2\rho}} G _ { 1,3 } ^ { 3,0 } \left( \frac { \vartheta^2 } { 4 }\bigg| \begin{array} { c } 0 \\ - 1  , \frac{m_2L-B_1}{2}+\rho,  -\frac{m_2L-B_1}{2}+\rho \end{array} \right).
 \end{aligned}
\end{array}
\end{equation}
\hrule
\end{scriptsize}
\end{table*}
\vspace{-0.1cm}
%\iffalse
 \begin{figure*}
\addtocounter{figure}{-1}
\subfigure{
\begin{minipage}[t]{0.3\linewidth}
\vspace{-1em}
  \centering
{\includegraphics[scale=0.3]{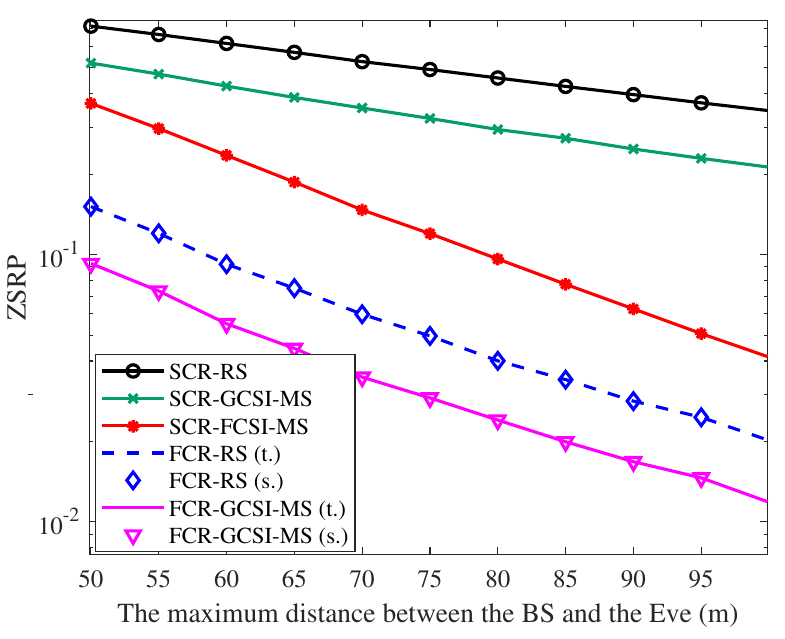}\\
  \linespread{0.85}{\caption{The ZSRP versus the maximum distance between the UAV-BS and the Eve, where “t.” and “s.” stand for theoretical and simulation results, respectively.}\label{Fig2}}}
  \end{minipage}
}
\quad
\subfigure{
\begin{minipage}[t]{0.3\linewidth}
\vspace{-1em}
  \centering
{\includegraphics[scale=0.3]{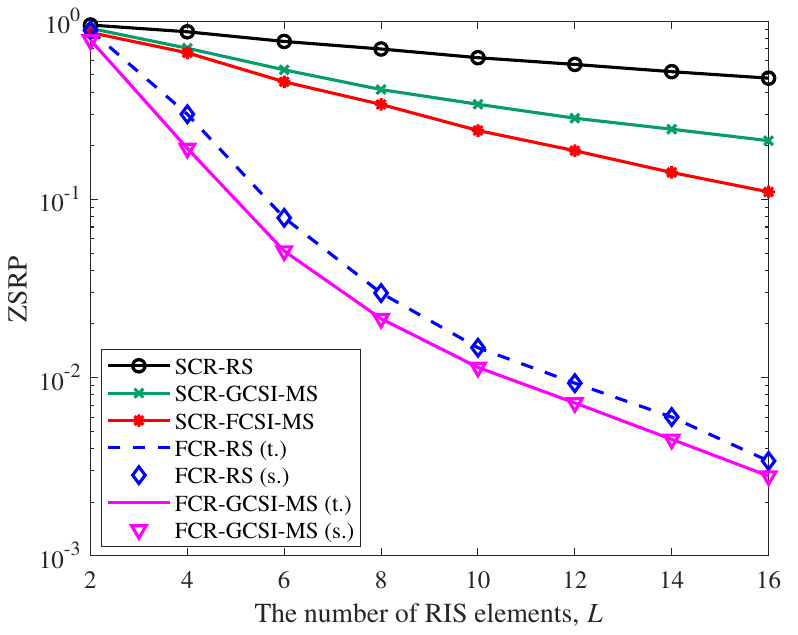}\\
  \linespread{0.85}{\caption{The ZSRP versus the number of RIS elements.}\label{Fig3}}}
  \end{minipage}
}
\quad
\subfigure{
\begin{minipage}[t]{0.3\linewidth}
\vspace{-1em}
  \centering
{\includegraphics[scale=0.3]{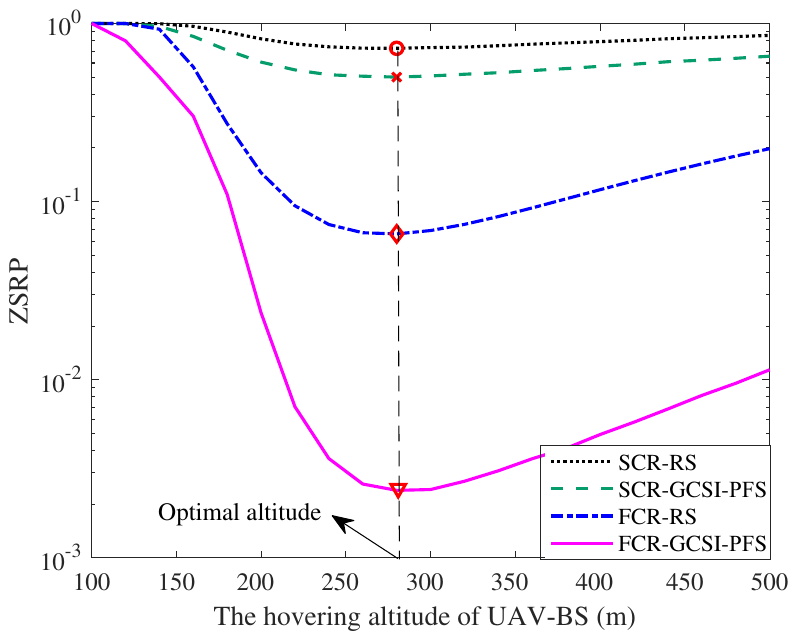}\\
  \linespread{0.85}{\caption{The ZSRP versus the hovering altitude of UAV-BS.}\label{Fig4}}}
\end{minipage}
}
\end{figure*}
%\fi
\vspace{-0.3cm}
\section{Numerical Results and Discussions}
\vspace{-0.2cm}
 In this section, we present simulation results to highlight the performance gains achieved by the FCR-RS and FCR-GCSI-PFS schemes as well as to verify the closed-form analysis. Moreover, the Monte-Carlo simulation demonstrates the GCSI-based multiuser scheduling is a globally optimal strategy in FC-RIS-assisted communications. Unless otherwise stated, numerical parameters are specified as follows. The radius of possible locations of the Eve is 500 m, which is the maximum distance between the UAV-BS and the Eve. Also, the distances between the UAV-BS and the FC-RIS, between the FC-RIS and the $n$-th user are 300 m and 50 m, respectively denoted by $d_{\text{BR}}$ and $d_{\text{R}n}$. Besides, the number of RIS elements is 16, and the ratio between the transmit power at the UAV-BS and the noise is 20 dB. All Nakagami fading parameters are set as $m_k=2$, $k\in\{{\text{R}n},{\text{BR}}\}$. For comparison purposes, we introduce the benchmark schemes below.

 1) SCR-RS: This scheme considers conventional single-connected (SC)-RIS-assisted communications where multiple users are orderly scheduled, serving as a counterpart to the FCR-RS scheme.

 2) SCR-GCSI-PFS: In this scheme, multiple users are selected as the PFS by comparing the channel quality of links from the users to the RIS, and a conventional RIS is introduced as well. It serves as a counterpart of the FCR-GCSI-PFS scheme.

 3) SCR-FCSI-PFS: In this scheme, multiple users are selected as the PFS by comparing the channel quality of the cascaded links. It achieves the best performance utilizing perfect knowledge of all channels with a conventional RIS.

 From Fig. 2, we observe the ZSRP performance comparison of the FCR-RS and FCR-GCSI-PFS schemes, with solid lines (theoretical expressions) and discrete markers (Monte-Carlo simulations), whose well agreements verify our closed-form ZSRP analysis. Thus, the obtained expressions not only provide an efficient alternative to time-consuming simulations for evaluating secrecy performance, but also validate the full utilization of CSI in the FCR-GCSI-PFS scheme. This is associated with the fact that for each user, the link from the UAV BS to the RIS establishes shared channels. Hence, the channel quality differences of each user stem from the link from the RIS to the users. If each single-hop CSI of the cascaded channels is obtained, then the GCSI-based scheduling significantly reduces the overhead to acquire CSI. When the UAV BS gets away from the Eve, the ZSRPs of all schemes decrease, indicating more secure communications. Additionally, the performance superiority achieved by the FCR-GCSI-PFS or SCR-FCSI-PFS scheme compared with the SCR-GCSI-PFS is accompanied with a higher complexity for synchronization and optimization implementation of FC-RIS.

 Fig. 3 depicts the ZSRP of the FCR-RS, FCR-GCSI-PFS, and SCR-RS, SCR-GCSI-PFS, SCR-FCSI-PFS schemes versus the number of RIS elements. With an increasing number of RIS elements, the ZSRPs of all schemes decrease accordingly. Notably, employing a FC-RIS obviously reduces the ZSRPs, compared with adopting a conventional RIS. It can be seen from Fig. 3 that the ZSRP of the SCR-FCSI-PFS scheme is much worse than that of either FCR-RS or FCR-GCSI-MS when a FC-RIS is introduced.

 Fig. 4 shows the ZSRP of the FCR-RS, FCR-GCSI-PFS, and SCR-RS, SCR-GCSI-PFS, SCR-FCSI-PFS schemes versus the hovering altitude of the UAV-BS. With an increasing UAV altitude, the ZSRPs of all schemes initially decrease and then start to increase with a substantially high altitude, implying the existence of an optimal hovering altitude of the UAV-BS. This existence is owing to the adopted aerial channel model consisting of probabilistic LoS links with altitude-variant path loss exponents. To determine a desired 3D location ${\bf{B}}$, some heuristic algorithms (e.g., golden search, particle swarm optimization, etc.) can be applied with the closed-form expressions. It can be observed from Fig. 4 that the optimal hovering altitude of the UAV-BS with a FC-RIS is the same as that considering a conventional SC-RIS. This is because that the distance-dependent large-scale fadings of the FC-RIS-aided communications have identical features with the SC-RIS-aided communications.
\vspace{-0.5cm}
\section{Conclusion}%\linespread{0.9}{
In this paper, we studied the PLS of FC-RIS and UAV empowered multiuser communications. We proposed the FCR-RS and FCR-GCSI-PFS schemes to eliminate the need for the rapidly-varying CSI brought by the high mobility of the UAV. A key observation through simulations was that to secure multiuser communications of interest, different multiuser scheduling designs preferred to the same optimal hovering altitude of the UAV-BS no matter whether a FC-RIS or a SC-RIS was introduced.%}
\linespread{0.6}
\vspace{-0.2cm}
\begin{tiny}

\end{tiny}

\clearpage

\end{document}